\documentstyle[amssymb,12pt,graphics,aps]{revtex}

\begin{document}
\draft
\title{Integrability in 1D Quantum Chaos}
\author{Yu. Dabaghian, R. V. Jensen and R. Bl\"umel}
\address{Department of Physics, 
Wesleyan University,\\
Middletown, CT 06459-0155}
\date{\today}
\maketitle

\begin{abstract}
Explicit, exact periodic orbit expansions for individual eigenvalues exist 
for a subclass of quantum networks called regular quantum graphs. We prove
that all linear chain graphs have a regular regime.
\end{abstract}

\section{Quantum Networks}

As a model of a quantum chaotic system, let us consider a quantum particle 
that moves on a quasi one-dimensional network. On every bond $B_{ij}$ of 
the network one may define a potential $U_{ij}$, which turns the graph 
into a {\em dressed} quantum graph \cite{Nova}.
Below we shall consider the case of constant {\em scaled} bond potentials 
$U_{ij}=E\lambda_{ij}$, where $E$ is the energy and $0<\lambda_{ij}<1$ are
real scaling constants.
Such potentials will change the action lengths of the bonds and the values 
of the reflection and transmission amplitudes of the quantum particle at 
the vertices of the graph. The wave function of the particle is given by
\begin{equation}
\Psi_{ij}(x)=A_{ij}e^{ik_{ij}x_{ij}}+B_{ij}e^{-ik_{ij}x_{ij}}.
\label{psi}
\end{equation}
The scaling assumption allows to avoid unnecessary complications, such 
as the geometrical restructuring of the periodic orbits for different
energies.

The boundary conditions consist of continuity and flux conservation
requirements for the wave function $\Psi_{ij}$ across the vertices $V_{i}$ 
of the graph \cite{Nova,Opus,Prima,Sutra,Stanza,Maxima}.
Such boundary conditions produce a spectral determinant $\Delta(k)$, 
whose zeroes define the spectrum of the problem. 
Usually, the quantum spectrum
determines the energy levels for a given potential. However, for the
scaling problems considered here, the scaling constants are given and the
potential is determined by the spectrum.

The spectral determinant for the quantum networks is a finite 
exponential sum,
\begin{eqnarray}
\Delta(k)=1+e^{i2(S_{0}k-\pi\gamma_{0})}
-\sum_{i=1}^{2N_{\Gamma}}a_{i}e^{i2(S_{i}k-\pi\gamma_{i})},
\label{polyn}
\end{eqnarray}
where $S_{i}$ are constants that can be expressed via the action lengths 
of the graph bonds and the $\gamma_{i}$ are constants. 
The total length $S_{0}$ (the 1D volume) of the graph is the largest 
frequency that appears in the exponential sum (\ref{polyn}), $S_{0}>S_{i}$. 
The number of terms $N_{\Gamma}$ in the sum (\ref{polyn}) is finite.

We are interested in studying the roots of $\Delta(k)$. Such functions 
(for a more general case of a complex argument) were studied from the 
mathematical perspective in \cite{Levin}, where it was shown that the 
roots $z_{n}$ of (\ref{polyn}) form an almost periodic set, which is 
uniformly bounded in the complex plane, i.e. is contained in a stripe 
of finite width,
\begin{equation}
\left|\mathop{\rm Im} z_{n}\right|<M<\infty, \ \ \Delta(z_{n})=0.
\label{zeropolyn}
\end{equation}
Moreover, it was shown that the real parts of the roots have the form
\begin{equation}
\mathop{\rm Re} z_{n}=\frac{\pi}{S_{0}}n+\varphi(n),
\label{realpart}
\end{equation}
where $\varphi(n)$ is an almost periodic function, whose frequency set 
contains linear combinations of the frequencies $\pi S_{i}/S_{0}$, 
with integer coefficients.
In \cite{Nova,Opus,Prima,Sutra,Stanza} these mathematical results were 
put into the perspective of quantum chaos theory that describes the 
spectral properties of quantum networks.

\section{Quantum Chaos}

The spectral problem of quantum networks has a long history (see 
\cite{Roth,QGT,Orsay} and the references therein).
In fact, the problem of obtaining the energy levels in the step potential
\begin{equation}
V(x) =\cases{
0, &for $0<x\leq b$, \cr
V_0=\lambda E, &for $b<x<1$,\cr
\infty, &for $ x<0$ or $x> L$,\cr
}
\label{pot}
\end{equation}
is the one that is usually presented following the elementary ``infinite 
square well'' problem in many standard quantum mechanics textbooks 
\cite{LL,Messiah,Flugge}.
However, unlike in the square well problem, the spectral equation for the 
step potential (\ref{pot}) is not easy to solve analytically. In fact, the 
only known way to obtain the roots of the spectral equation are numerical 
or graphical solutions.

The complexity of the spectral equation for the system (\ref{pot}) as well
as for a generic quantum network is in fact due to the dynamical complexity 
of the underlying classical system, which turns out to be chaotic \cite{Gaspard}.

In the classical limit the quantum graph systems reduce to a classical 
particle that scatters stochastically at the vertices of the graph (Fig.~3).
The scattering probabilities are obtained in the $\hbar=0$ limit from 
the quantum-mechanical vertex transition amplitudes $t_{ji,ij}$ \cite{Stanza}.
Every time the particle encounters a vertex $V_{i}$ traveling along the bond 
$B_{ji}$, it can reflect back from it with the probability $|t_{ji,ij}|^{2}$ 
or pass through to another bond $B_{ik}$ with the transmission probability 
$|t_{ji,ik}|^{2}$.
After every reflection or transmission (scattering) event, the particle 
completely loses its memory about the previous stage of its motion. 

Using the coefficients $t_{ji,ik}$, it is easy to obtain the probability
amplitudes with which a particular periodic orbit is realized. Indeed, if a 
certain orbit $\gamma$ reflects $\sigma_{i}(\gamma)$ times off the vertex 
$V_{i}$ and passes through it $\tau_{i}(\gamma)$ times, then the probability 
amplitude for such an orbit is
\begin{equation}
A_{\gamma}= \prod t_{ji,ij}^{\sigma_{i}(\gamma)}t_{ji,ik}^{\tau_{i}(\gamma)},
\label{A}
\end{equation}
where the product is taken over all the vertices visited by $\gamma$.
It is usually convenient to distinguish the so-called ``primitive'' orbits,
i.e. the ones that traverse a certain sequence of bonds only once, and the
``multiple traversal'' orbits.
If a certain primitive orbit $\gamma_{p}$ with the weight $A_{p}$ is traversed 
$\nu$ times, then the corresponding amplitude is $A_{p}^{\nu}$.

Due to the stochastic scattering, the trajectories of the particle are 
geometrically very complex and the number of possible orbits (and in 
particular the periodic orbits), increases exponentially \cite{Gaspard}.
The number $N$ of the $n$-bond periodic orbits grows as
\begin{equation}
N\approx \frac{e^{\tau n}}{n},
\label{rate}
\end{equation}
where the exponential proliferation rate $\tau$ depends only on the topology 
of the graph.

Such behavior mimics closely an important phase space feature of 
deterministically chaotic systems. Hence quantum networks provide very 
convenient and simple models for studying various features of quantum 
systems that correspond to classically chaotic systems in the context 
of quantum chaos theory.

\section{Periodic orbit theory}

Let us outline the main attributes of quantum chaos theory \cite{Gutzw}, 
applied to quantum networks.
In \cite{Nova} it was shown that for dressed quantum graphs there 
exists an exact periodic orbit expression for the density of states, 
\begin{eqnarray}
\rho(k)=\frac{S_{0}}{\pi}+\frac{1}{\pi}\mathop{\rm Re}\sum_{p}
S_{p}^{0}\sum_{\nu=1}^{\infty}
A_{p}^{\nu}e^{i\pi\nu S_{p}^{0}k},
\label{rho}
\end{eqnarray}
which generalizes the earlier results obtained in \cite{Roth,QGT}. 
Here $S_{p}^{0}$ and $A_{p}(E)$ are correspondingly the action length and the 
weight factor (\ref{A}) of the prime periodic orbit labeled by $p$, $\nu$ 
is the repetition index and $S_{0}$ is the total action length of the graph.

Another important characteristic of the spectrum is the spectral staircase 
function
\begin{equation}
N(k)=\sum_{j}\Theta(k-k_{j}),
\label{staircase}
\end{equation}
for which there also exists an exact periodic orbit expansion.
In accordance with Weyl's law, the average spectral staircase depends linearly 
on the momentum,
\begin{equation}
\bar N(k)=\frac{S_{0} k}{\pi}+\bar N(0),
\label{weyl}
\end{equation}
where $S_{0}$ is the total length (1D volume) of the network.
This statement implies that the average dependence of the momentum 
eigenvalues $k_{n}$ on their index \cite{Stanza} is
\begin{equation}
\bar k_{n}=\frac{\pi}{S_{0}}\left(n-\bar N(0)-\frac{1}{2}\right),
\label{averoot}
\end{equation}
which complies with the relationship (\ref{realpart}).

Weyl's average (\ref{weyl}) of the spectral staircase is the same for all 
networks with the same action length $S_{0}$. However, different networks 
vary significantly in how well the average approximates the actual 
staircase function $N(k)$.
It is interesting that there exist certain graphs with a particularly 
regular behavior of their spectra. For these quantum networks (called 
regular in \cite{Opus,Prima,Sutra,Stanza}), the average spectral
staircase (\ref{weyl}) pierces every stair-step of $N(k)$.
It is clear that for the regular graphs the intersection points $\hat k_{n}$ 
between the spectral staircase and its average,
\begin{equation}
N(\hat k_{n})=\bar N(\hat k_{n}),
\label{separators}
\end{equation}
occur periodically on the momentum axis \cite{Thales}. 
In other words, for such systems there exists exactly one quantum eigenvalue 
$k_{n}$ between every two such intersections. 

For such graphs one can obtain immediately the exact periodic orbit series
representation for the {\em individual} eigenvalues of the momentum.
Indeed, since every interval $I_{n}=[\hat k_{n-1},\hat k_{n}]$ contains one 
delta-peak of $\rho(k)$,
the corresponding root $k_{n}$ can be obtained via 
\begin{equation}
k_{n}=\int^{\hat k_{n}}_{\hat k_{n-1}}k\rho(k)dk.
\label{eigenvalue}
\end{equation}
Using the periodic orbit expansion for the density of states (\ref{rho}),
the integration (\ref{eigenvalue}) can be performed explicitly and yields
\cite{Opus,Stanza}
\begin{eqnarray}
k_{n}=\frac{\pi}{S_{0}}n-\frac{2}{\pi}\sum_{p}
\frac{1}{S_{p}^{0}}\sum_{\nu=1}^{\infty}
\frac{A_{p}^{\nu}}{\nu^{2}}\sin\left(\frac{1}{2}\nu\omega_{p}\right)\,
\sin(\nu\omega_{p}n),
\label{kn}
\end{eqnarray}
where $\omega_{p}=\pi S_{p}^{0}/S_0$, and the $A_{p}$'s are assumed to be real
(no mixed boundary conditions).

The spectral formula (\ref{kn}) provides an explicit harmonic expansion
of the function $\varphi(n)$ in (\ref{realpart}) for the roots of the 
spectral determinant of the regular graphs.

It can be shown \cite{Maxima} that the series (\ref{kn}) is convergent.
An analytical criterion for the regularity (\ref{separators}) of a given 
network comes from analyzing the spectral determinant (\ref{polyn}). 
After removing the complex phase of $\Delta(k)$, the spectral equation 
for quantum graphs can be written as 
$\cos\left(S_{0}k-\pi\gamma_0\right)=\Phi(k)$.
It was shown in \cite{Opus,Prima,Sutra,Stanza,Maxima}, that if the {\em 
characteristic function} $\Phi(k)$ of the graph is bounded by $1$ for all 
$k\in R^{1}$, then the piercing average condition (\ref{separators}) is 
satisfied \cite{Opus,Stanza} and the expansion (\ref{kn}) exists.
In terms of the coefficients of (\ref{polyn}), $|\Phi(k)|<1$ is certainly 
satisfied if
\begin{equation}
\sum_{i=1}^{N_{\Gamma}}|a_{i}|<1.
\label{regular}
\end{equation}
In \cite{Opus,Prima,Sutra,Stanza,Maxima} we called graphs satisfying
(\ref{regular}) regular quantum graphs.
In general, the regularity condition is satisfied only for a special 
choice of the potentials (hence the importance of using various ``graph
dressings'') \cite{Nova,Stanza}.

\section{Linear chain graphs}

For many network topologies there exist no regular regimes at all 
\cite{Stanza}. As an example of graphs which can be made regular 
by an appropriate choice of parameters, let us consider a dressed 
chain graph (a sequence of $1D$ square wells), with the vertices 
positioned at $x=b_{0},x=b_{1},...,x=b_{n}$ starting at $b_{0}=0$. 
For every region $b_{i-1}<x<b_{i}$, the wave function is given by
\begin{equation}
\psi _{i}(x)=\frac{A_{i}}{\sqrt{k_{i}}}e^{ik_{i}x}+
\frac{B_{i}}{\sqrt{k_{i}}}e^{-ik_{i}x}, 
\end{equation}
where $k_{i}=\sqrt{k^{2}-\lambda_{i}k^{2}}=\beta_{i}k$, and where the 
physically meaningful flux amplitudes $1/\sqrt{k_{i}}$ are factored 
out of the arbitrary coefficients $A_{i}$ and $B_{i}$.
The boundary conditions at every vertex, 
\begin{eqnarray*}
\frac{A_{i}}{\sqrt{\beta_{i}}}e^{ik\beta_{i}b_{i}}+
\frac{B_{i}}{\sqrt{\beta_{i}}}e^{-ik\beta_{i}b_{i}} 
&=&
\frac{A_{i+1}}{\sqrt{\beta_{i+1}}}e^{ik\beta_{i+1}b_{i}}+
\frac{B_{i+1}}{\sqrt{\beta_{i+1}}}e^{-ik\beta_{i+1}b_{i}}, \\
A_{i}\sqrt{\beta_{i}}e^{ik\beta_{i}b_{i}}-B_{i}\sqrt{\beta_{i}}e^{-ik\beta_{i}b_{i}} 
&=&
A_{i+1}\sqrt{\beta_{i+1}}e^{ik\beta_{i+1}b_{i}}-
B_{i+1}\sqrt{\beta_{i+1}}e^{-ik\beta_{i+1}b_{i}}
\end{eqnarray*}
can be written in the matrix form 
\begin{equation}
M_{i+1,i}\vec{C}_{i+1}=M_{i,i}\vec{C}_{i},
\end{equation}
where
\begin{equation}
M_{ij}\equiv \pmatrix{
{\frac{1}{\sqrt{\beta_{i}}}e}^{{ik\beta_{i}b_{j}}} & {\frac{1}{\sqrt{\beta
_{i}}}e}^{{-ik\beta_{i}b_{j}}} \cr 
{\sqrt{\beta_{i}}e}^{{ik\beta_{i}b_{j}}} & -{\sqrt{\beta_{i}}e}^{-{ik\beta_{i}b_{j}}}
} ,\ \ \ 
\vec{C}_{i}\equiv 
\pmatrix{
{A_{i}} \cr 
{B_{i}}
}.
\end{equation}
The amplitudes $\vec{C}_{2}$ at a vertex $V_{i}$ can be expressed via the amplitudes 
at the previous vertex by means of the transfer matrix $T_{i}$,
\begin{equation}
\vec{C}_{i+1}=M^{-1}_{i+1,i}M_{i,i}\vec{C}_{i}\equiv T_{i}\vec{C}_{i},
\label{match}
\end{equation}
where
\begin{equation}
T_{i}=\frac{1}{t_{i}}
\pmatrix{
e^{ik\left( \beta_{i}-\beta_{i+1}\right) b_{i}} & 
r_{i}e^{-ik\left(\beta_{i+1}+\beta_{i}\right) b_{i}} \cr
r_{i}e^{ik\left( \beta_{i+1}+\beta_{i}\right) b_{i}} & 
e^{ik\left( \beta_{i+1}-\beta_{i}\right) b_{i}}
}
\label{bond}
\end{equation}
is unitary, $\det T_{i}=1$.
Here $r_{i}$ is the reflection and $t_{i}=\sqrt{1-r_{i}^{2}}$ is the
transmission coefficient, 
\begin{eqnarray*}
r_{i} &=&\frac{\beta_{i+1}-\beta_{i}}{\beta_{i+1}+\beta_{i}}, \cr
t_{i} &=&\frac{2\sqrt{\beta_{i}\beta_{i+1}}}{\beta_{i}+\beta_{i+1}}.
\end{eqnarray*}
Applying the ``boundary matching'' equation (\ref{match}) consecutively
to all vertices of the chain yields the equation 
\begin{equation}
\vec{C}_{n}=T\vec{C}_{1},  
\label{n1}
\end{equation}
where 
\begin{equation}
T=T_{n-1}T_{n-2}...T_{1}.
\label{totalT}
\end{equation}
Due to the Dirichlet boundary conditions at the end vertices $V_{0}$ and 
$V_{n}$ of the chain,
\begin{equation}
\vec{C}_{1}=\pmatrix{
A{_{1}} \cr
-A{_{1}}
} ,\quad 
\vec{C}_{n}=\pmatrix{
A{_{n}} e^{-ik_{n}b_{n}} \cr 
-A{_{n}}e^{ik_{n}b_{n}}
},
\label{dirichlet}
\end{equation}
equation (\ref{n1}) can be written in terms of the 
matrix elements of $T$ only,
\begin{equation}
-e^{ik_{n}b_{n}}\left[t_{11}(k)-t_{12}(k)\right]=
e^{-ik_{n}b_{n}}\left[t_{21}(k)-t_{22}(k)\right].  
\label{entries}
\end{equation}
Since the spectral equation (\ref{entries}) is linear, the common factor
\begin{equation}
\tau^{-1}=\prod_{i=1}^{n}t_{i}^{-1}  
\label{tr}
\end{equation}
that multiplies the matrix $T$ (\ref{totalT}) cancels out, so that 
(\ref{entries}) contains only the reflection coefficients $r_{1}$, 
$r_{2}$,..., $r_{n}$, and no transmission coefficients $t_{i}$.
From the explicit form of $T_{i}$ it follows that the matrix elements 
$t_{ij}$ of $T$,
\[
\left(\prod_{i=1}^{n}\frac{1}{t_{i}}\right) \pmatrix{
e^{-ik_{n}b_{n}}e^{ikS_{0}}+e^{-ik_{n}b_{n}}P\left( r_{i},k\right) & 
e^{-ik_{n}b_{n}}Q\left( r_{i},k\right) \cr
e^{ik_{n}b_{n}}Q^{*}\left( r_{i},k\right) & 
e^{ik_{n}b_{n}}e^{-ikS_{0}}+e^{ik_{n}b_{n}}P^{*}\left(r_{i},k\right)
},
\]
are certain multivariable polynomials of the variables $r_{i}$ that 
may include monomials of all the degrees between $1$ and $n$. Hence 
the coefficients $a_{i}$ in the spectral determinant 
\begin{eqnarray}
\Delta(k) =
e^{2ik_{n}b_{n}}\left(t_{11}(k)-t_{12}(k)\right)+t_{21}(k)-t_{22}(k)
\cr
=1+e^{i2(S_{0}k-\pi\gamma_{0})}
-\sum_{i=1}^{2N_{\Gamma}}a_{i}e^{i2(S_{i}k-\pi\gamma_{i})},
\label{determinant}
\end{eqnarray}
are also polynomials in $r_{i}$ of degree larger or equal to $1$.

Hence, by choosing the graph dressing parameters $\beta_{i}$
sufficiently close to one another, one can make the coefficients $r_{i}$
as small as necessary in order to satisfy the regularity condition 
(\ref{regular}). Hence the regular regime (\ref{separators}) exists 
for all the linear chain graphs. Physically this corresponds to the 
case where the heights of the bond potentials do not differ too much 
from one another.

As an example, consider the case of a two-bond linear graph with Dirichlet
boundary conditions at the end vertices positioned at $x=0$ and $x=b_{2}$ 
\cite{Opus,Prima,Sutra,Stanza}. In this case the transfer matrix is
\begin{equation}
T=\frac{1}{t}\pmatrix{
   e^{ik\left(\beta_{1}-\beta_{2}\right)b_{1} } & 
r\,e^{-ik\left(\beta_{1}+\beta_{2}\right)b_{1} } \cr 
r\,e^{ik\left(\beta_{1}+\beta_{2}\right)b_{1} } &
e^{ik\left(\beta_{2}-\beta _{1}\right)b_{1} }
},
\label{twohydra}
\end{equation}
where $r$ and $t$ are correspondingly the reflection and the transmission 
coefficients at the middle vertex $V_{2}$.
The spectral equation obtained from (\ref{twohydra}) is
\begin{eqnarray}
\sin \left[k\left(\beta_{1}b_{1}+\beta_{2}\left(b_{2}-b_{1}\right)\right)\right]
+
r\sin \left[k\left(\beta_{1}b_{1}-\beta_{2}\left(b_{2}-b_{1}\right)\right)\right] 
=0.
\end{eqnarray}
Since $r\leq 1$, the condition (\ref{regular}) is satisfied and 
therefore a two-bond chain graph is always regular \cite{Opus,Prima,Sutra,Stanza}.

\section{Irregular regimes}
As a network moves from the regular to the irregular regime, the principle
of ``one root per interval $I_{n}$'' is violated. In the irregular regime 
certain intervals contain no roots of the spectral equation at all, while 
others may host two or more roots. The piercing property of the Weyl average 
\cite{Opus,Stanza} also disappears, so in order to separate the intervals 
that contain exactly one root, one cannot use Weyl's average and the 
intersection points (\ref{separators}).

As outlined in \cite{Stanza}, in order to repeat the root expansion
procedure discussed above, one would need to find some smooth monotonic
function 
$\bar N_{1}(k)$, which pierces every stair of $N(k)$, and find
the intersection points $\bar N_{1}(\hat k'_{n})=N(\hat k'_{n})$.
Solving the equation $\bar N_{1}(\hat k'_{n})=n$ would lead to the 
``generalized separators'' $\hat k'_{n}$, that isolate the roots 
$k_{n}$ from each other.
Unlike in the regular regime, where the separators $\hat k'_{n}$ are 
periodic, the separators $\hat k'_{n}$ form an almost periodic set \cite{Stanza}.

\section{Discussion}

Since quantum chaos theory deals with non-integrability in the
semiclassical regime, the systems it usually addresses are fairly 
complicated.
One of the drawbacks of such complexity is that the various methods 
used in the theory (such as periodic orbit theory, random matrix 
theory, etc.) describe the system from quite different perspectives, 
and the direct connections between these methods are often obscured.

Quantum networks provide a convenient exception to this.
For these systems, many essential attributes of quantum chaos 
theory, and in particular the periodic orbit theory, can be directly 
established, and the results turn out to be exact.
As a result, quantum networks provide a convenient model for testing 
various phenomenological and approximate methods.

Due to their simplicity, analytical solvability and overall transparency,
quantum graphs play the role of the ``harmonic oscillator'' of the 
quantum chaos theory.

\section*{Acknowledgments}

Y.D. and R.B. gratefully acknowledge financial support by NSF grant 
PHY-9984075; Y.D. and R.V.J by NSF grant PHY-9900746.


\end{document}